\documentstyle[preprint,eqsecnum,prd,aps]{revtex}
\def\frc#1#2{{\textstyle{#1 \over #2}}}

\def\prd#1#2#3{, {\it Phys.\ Rev.\ D} {\bf #1}, #2 (19#3)}
\def\plb#1#2#3{, {\it Phys.\ Lett.\ B} {\bf #1}, #2 (19#3)}
\begin{document}

\draft
\preprint{hep-ph/9507345 \hbox{\hskip3.9in} HUTP-95/A025}

\title{Pseudoscalar Conversion and X-rays from the Sun}

\author{Eric D. Carlson\footnote{Address after August 1: Olin Physical
Laboratory, Wake Forest University, Winston-Salem NC  27109}}
\address{Lyman Laboratory of Physics, Harvard University \\
 Cambridge, Massachusetts 02138}

\author{Li-Sheng Tseng}
\address{Lyman Laboratory of Physics, Harvard University \\
 Cambridge, Massachusetts 02138}

\date{July 18, 1995}

\maketitle

\bigskip\bigskip

\centerline{\bf Abstract}

We investigate the detection of a pseudoscalar $\phi$ that couples
electromagnetically via an interaction ${1\over4}g \phi F {\tilde F}$.  In
particular, we focus on the conversion of pseudoscalars produced in the
sun's interior in the presence of the sun's external magnetic dipole field
and sunspot-related magnetic fields.  We find that the
sunspot approach is superior.  Measurements by the SXT on the Yohkoh
satellite can measure the coupling constant down to $g=0.5$--$1 \times
10^{-10}\,\rm GeV^{-1}$, provided the pseudoscalar mass $m < 7{\times}
10^{-6}\,$eV, which makes it competitive with other astrophysical
approaches.

\vfill\eject

\section{Introduction}

Light neutral pseudoscalars arise naturally as a result of spontaneously
broken global symmetries \cite{kim}.  They will be truly massless if the
symmetry is not anomalous.  Although some models can be directly tested in
colliders, some of the best limits come from astrophysics.

Such particles generically couple to the electromagnetic field through
couplings of the form
\begin{equation}
	{\cal L}= \frc14 g \phi F^{\mu\nu}{\tilde F_{\mu\nu}} \; .
	\label{ffdual}
\end{equation}
where $\tilde F_{\mu\nu}$ is the dual of the electromagnetic field
$F^{\mu\nu}$ and $g$ is a coupling constant.  This coupling can lead to the
interconversion of photons and pseudoscalars in a background
electric or magnetic field.

It is reasonable to attempt to use such a coupling to attempt to detect or
exclude pseudoscalars of a given mass or coupling.  There has been a recent
resurgence of interest in detection using such couplings.  Several
astrophysical approaches have been considered for detecting these
electromagnetic couplings \cite{allsearch}, including penetration of light
from distant stars through intervening opaque clouds \cite{garretson}, the
effect on evolution of stars \cite{hbstars,pmass}, and the appearance of
X-rays generated in the cores of distant stars that have been double
converted to pseudoscalars and back to photons again \cite{carlson}.  In
all but the first of these methods, the pseudoscalars are generated from
thermal photons in the presence of nuclear electric fields.

The sun is the nearest and hence relatively brightest source of
pseudoscalars.  Naturally it has been the source of some recent
pseudoscalars search proposals and experiments.  Lazarus
and his collaborators searched for solar pseudoscalars by
attempting to detect a signal from pseudoscalar conversion in a static
magnetic field on the earth \cite{search}.  The probability of
pseudoscalar conversion in such a helioscope is proportional to $g^2 B^2
\ell^2$.  For this experiment, where the magnetic field is
$B=2.2{\times}10^4\,$gauss and its length $\ell =1.8\,$m, failure to detect
a signal set the limit at $g=3.6 {\times} 10^{-9}\,\rm GeV^{-1}$, provided
the pseudoscalar mass $m < 0.03\,$eV.  A more expansive search is taking
place in Novosibirsk, where a gimballed accelerator dipole magnet tracks
the sun \cite{gimbal}.  With $B=6{\times}10^4\,$gauss and $\ell =6\,$m, the
detector should be able to provide sensitivity down to about
$g=10^{-10}\,\rm GeV^{-1}$, provided $m < 0.1\,$eV.  Given that maintaining
a much stronger and larger magnetic field than the one used in Novosibirsk
may not be possible, the limit from the Novosibirsk search may be the best
attainable from this method of pseudoscalar-photon conversion on earth.

There are, however, other sources of magnetic field in which solar
pseudoscalars can convert to photons.  On the surface of the sun, there
exist many highly dynamic magnetic structures.  Pseudoscalars emerging from
the sun's interior may convert to X-rays in the sun's own external magnetic
fields.  These X-rays should be detectable by solar X-ray telescope, such
as the SXT on the Yohkoh satellite.

Two strong candidates for the source of pseudoscalar conversion in the
sun's exterior are the dipole field and sunspot-related fields.  The sun's
dipole field, although rather weak, has an effective length of a few hundred
thousand kilometers.  Sunspots have field strengths of a few kilogauss on the
sun's surface with an effective length of no less than a few thousand
kilometers.  We find that the magnetic field of the sunspot produces a
significantly larger X-ray count rate than the sun's general dipole field.
Furthermore, using the magnetic field of a large sunspot, it is possible
with Yohkoh's SXT to set the limit $g<0.5$--$1 \times 10^{-10}\,\rm
GeV^{-1}$, provided  $m < 7{\times} 10^{-6}\,$eV if no X-ray signature from
pseudoscalar conversion is detected.  This makes this method competitive
with other approaches.  However, the region we are able to exclude is
almost completely covered by the horizontal branch star approach, which
covers a higher mass range and couplings down to $6{\times}10^{-11}\,\rm
GeV^{-1}$.  Because of the importance of redundant checking in
astrophysics, and the relative ease of completing this analysis, we feel
this approach is worthy of further consideration.

In section 2 we discuss the production of pseudoscalars from the sun's
interior.  In section 3 we discuss their conversion to photons in the
presence of the sun's external magnetic fields and the limit which such
conversion can achieve.  Detection of converted photons by Yohkoh's
SXT is discussed in section 4.  We summarize our findings in section 5.

\eject

\section{Production of Pseudoscalars in the Sun}

Pseudoscalars with a coupling of the form (\ref{ffdual}) are produced
primarily by the Primakoff process.  In vacuum the cross section for this
process is logarithmically infinite, due to the infinite range of the
Coulomb potential.  However, at finite temperature and in a dense medium
such as the sun's, Debye screening cuts off this divergence and leads to
a conversion rate per unit time for photon to pseudoscalar of
\cite{primakoff}
\begin{equation}
      \Gamma = {g^2 \kappa^2 T \over 32 \pi}
         \left[ \left( 1 + {\kappa^2 \over 4E^2} \right)
         \ln \left(1 + {4E^2 \over \kappa^2} \right) -1 \right] \; ,
         \label{primak}
\end{equation}
where $E$ is the energy of a photon, $T$ is the temperature, and $\kappa$
is the inverse Debye screening length.  Noting that all except heavy
elements, which constitute less than 1\% of the total mass fraction of
sun's interior, dissociate completely, we shall assume for simplicity the
complete dissociation of all elements in the sun which gives
\begin{equation}
       \kappa^2 = {4 \pi \alpha \over T} \sum_i n_i \left( Z_i^2+Z_i\right)
       \; , \label{Debye}
\end{equation}
where $n_i$ and $Z_i$ are the number density and charge of each component
of the local density, respectively, and $\alpha$ is the fine structure
constant.  Multiplying (\ref{primak}) times the density of thermal photons,
we find the pseudoscalar production density per unit energy per second is
given by
\begin{equation}
       {d \Gamma \over dV \, dE} = {g^2\xi^2 T^3 \over 8\pi^3(e^{E/T}-1)}
       \left[(E^2 + \xi^2 T^2)\ln (1 + E^2/\xi^2 T^2) - E^2\right] \; ,
\end{equation}
where $\xi^2 = \kappa^2/4T^2$.

We now integrate this over a solar model, for which we use the
model of Bahcall and Ulrich \cite{model}.  The values of $T$ and
$\xi^2$ as functions of radius are graphed in Fig.\ 1.  The nearness of the
sun enables us to distinguish pseudoscalars emerging from different
points on the sun's disk.  This suggests that we should not
integrate pseudoscalar production throughout the volume of the sun, but
only along a line of sight.  Since the earth-sun distance is much greater
than the sun's radius, we shall treat the lines of sight as parallel.  The
total rate of production along a line of sight is thus given by
\begin{equation}
       {d \Gamma \over dx\,dy\, dE} ={g^2 R_\odot \over 4 \pi^3} F(E,b)\;,
\label{I}
\end{equation}
where the function $F(E,b)$ is defined as
\begin{equation}
       F(E,b)= \int_0^{\sqrt{1-b^2}} dz \; \left\{{\xi^2 T^3 \over
       (e^{E/T}-1)} \left[(E^2 + \xi^2 T^2)\ln (1 + E^2/\xi^2 T^2)
        - E^2\right] \right\}_{r=R_\odot \sqrt{z^2+b^2}}\;,
\end{equation}
where $\xi^2$ and $T$ are implicit functions of $r$, the distance from the
center of the sun, and $b$ is defined so that $bR_\odot$ is the impact
parameter between the line of sight and the center of the sun.  Values of
$F(E,b)$ for different energies at various $b$ are graphed in Fig.\ 2.

Note that most of the pseudoscalars are produced at $b < 0.2$.  Because
this is small, pseudoscalars emerge almost perpendicular to the sun's
surface.  In subsequent calculations, we will treat the sun's surface as flat,
and the pseudoscalars as emerging directly along a perpendicular. This
leads to errors proportional to $b^2$, or only a few percent.  Once the
pseudoscalars emerge from the surface of the sun, they can interact with
the sun's exterior magnetic field and back-convert into X-rays.   We now
proceed to find pseudoscalars' contributions to solar X-rays.

\section{Conversion of Pseudoscalars in the Sun's Magnetic Field}

In the presence of the sun's exterior magnetic field, pseudoscalars can
be converted into photons.  The probability of conversion along a
line of sight is given by \cite{garretson}
\begin{equation}
       P = \frc14g^2 \left\vert \vec D(x,y) \right\vert ^2\;,
\label{P}
\end{equation}
where we have introduced $\vec D(x,y)$ defined as
\begin{equation}
       \vec D(x,y) = \int_0^L {\vec B_\perp (x,y,z)}\,e^{i \theta(z)} \,
                     dz\;,
\end{equation}
where we have taken the line of sight to be the $z$ direction, $z=0$ being
where the pseudoscalars leave the sun, and $z=L$ being the earth.  The
phase $\theta(z)$ has two sources: the mass of the pseudoscalar (if any)
and the effective plasma mass of the photon due to the presence of plasma
outside the sun.  In general $\theta(z)$ will be given by
\begin{equation}
        \theta(z) = \int_0^z \left({2 \pi \alpha n_e(z') \over m_e E}
                  - {m^2 \over 2E} \right) \, dz' \; .
        \label{phase}
\end{equation}
where $m_e$ is the mass of the electron and $n_e$ is the total number
density of electrons, and $m$ is the mass of the pseudoscalar.

We will find later that there is only a significant effect if the
pseudoscalar is very light.  Let us assume for the moment that the
pseudoscalar is massless or so light that its contribution to the phase is
negligible.  The rapidly falling nature of the sun's coronal density
outside the chromosphere assures that the phase will oscillate rapidly
close to the sun and not change at all away from the sun.  The rapid
oscillations will wipe out the contribution from the magnetic fields very
near the sun.  For simplicity, we assume that the contribution is zero from
this nearby zone, and that the phase $\theta(z)$ is constant outside this
zone.  We define the boundary height $h$ by the somewhat arbitrary
condition $\frc\pi2 = \theta(\infty) - \theta(h)$.  Using numerical
models \cite{chrommodel} for the sun's chromosphere and corona, we find
$h=1200\,$km above the ``surface'' of the sun, defined as the one optical
depth level of the sun at a wavelength of $500\,$nm.  We can replace the
upper limit on integration by $\infty$, because of the short range of the
solar magnetic fields, so our formula for $\vec D$ is modified to
\begin{equation}
       \vec D(x,y) = \int_h^\infty {\vec B_\perp (x,y,z)}\, dz\;,
\label{D}
\end{equation}

We then integrate the production rate (\ref{I}) and conversion probability
(\ref{P}) over the regions of interest on the sun's surface
and divide by the solid angle of the sun to the earth to give
\begin{equation}
      {d\Phi \over dE} = {1 \over 4\pi L^2} \int \,dx \int \,dy
             {d \Gamma \over dE \, dx\,dy} P
       = {g^4 R_\odot \over 64 \pi^4 L^2} \int dx \, dy \,
            F(E,b) \left| \vec D(x,y) \right|^2 \; .
\label{counts}
\end{equation}
the differential X-ray flux at earth due to solar pseudoscalar-photon
coupling, where $b$ is an implicit function of $x$ and $y$.  This general
equation will be used to calculate the expected X-ray count rate coming
from two components of sun's exterior magnetic field; namely, the general
dipole field, and sunspots.

The sun's general field is most apparent during the periods of sunspot
minimum.  During these periods, magnetographs measure a weak field of
about one gauss around the poles \cite{solaract}.  Evidence that these
polar fields have opposite polarities have led to the belief that this
general field is a dipole.  Although the solar dynamo model predicts the
22-year cycle of the solar dipole moment \cite{babcock}, the exact behavior
of the general field is still uncertain.  For example, in
1957, just two years after the sunspot minimum, it was detected that the
two poles had the same polarity for a period of 18 months.  Nevertheless,
we shall assume the existence of the solar dipole field and calculate its
contribution to the X-ray count rate during periods of sunspot minimum.

The height $h$ where we can neglect plasma effects is small compared to the
scale of the dipole field, so to find $\vec D$ we begin integration at the
sun's surface.  Since the magnetic field of a dipole at the equator is half
the strength at the pole, and a dipole field falls as the cube of the
distance, we can find the value of $\vec D$ along our line of sight to the
center of the sun as $| \vec D_c| = \frc14B_pR_\odot$, where $B_p$ is the
magnetic field at the pole.  We have explicitly calculated $|\vec
D(x,y)|^2$ as a function of $x$ and $y$, and we find there is very small
variation over the region $b<0.2$ where pseudoscalar production is
important.  So we can approximate $|\vec D(x,y)| = |\vec D_c|$ everywhere,
and hence obtain with the help of (\ref{counts})
\begin{equation}
     {d\Phi_{\rm dip}\over dE} = {g^4 R_\odot^5 B_p^2\over 512 \pi^3 L^2}
                              \int_0^1 b F(E,b)\,db\;.
\label{dipcounts}
\end{equation}
Letting $g=10^{-10} {\rm GeV}^{-1}$ and $B_p=1\,$gauss, values of
(\ref{dipcounts}) for different energies are graphed in Fig.\ 3.  As
evident from Fig.\ 3, the sun's dipole field as the source of the magnetic
field for pseudoscalar conversion does not give a significant X-ray count
rate.  The signal would be detectable if
there were no background, but this, sadly, is not the case.  Fortunately,
sunspots are a much more promising source, and we now consider them.

As implied by (\ref{P}), strong magnetic field leads to a large pseudoscalar
to photon conversion probability.  On the sun's surface, the strongest
magnetic fields often arise from sunspots.  Beside having strong fields,
typically several kilogauss, other advantages of sunspots are
that they are visible on the sun's disk and that the X-ray background from
their central dark region, the umbra, and their surrounding less dark
region, the penumbra, is relatively small \cite{hudson}.  Although the
shapes of sunspots are often highly irregular, it is convenient to model
sunspots as having a circular umbra surrounded by an annular penumbra.  An
approximate field distribution for this model is given by a gaussian
\cite{spotfield},
\begin{equation}
      B_z(r) = {\phi \over \pi \sigma^2} \exp (-r^2/\sigma^2)\;,
\label{spotfield}
\end{equation}
where $B_z$ is the component vertical to the solar surface, $\phi$ is the
total magnetic flux coming out of the sunspot, $r$ is the radial distance
{}from the sunspot center, and $\sigma$ is the characteristic size of the
sunspot.  The radius of the penumbra $p$ is typically about $p^2=2.1
\sigma^2$.  Equation (\ref{spotfield}) is the field roughly at $z=0$.  For
the sunspots we are interested in, the size of the sunspot will be much
greater than $h=1200\,$km where we can neglect plasma effects on the
X-rays.  Presumably, the fields change relatively little over this
distance, and we will once again approximate $h=0$.

We now need to know the magnetic field, or, more to the point, $\vec
D(x,y)$ due to the sunspot.  Because the magnetic field is divergenceless,
and there is no magnetic field at infinity, we can show
\begin{equation}
           \nabla \cdot \vec D(x,y)  = B_z(x,y,z=0) \; .
\end{equation}
If we assume the magnetic field is rotationally symmetric above the surface
of the sun, this is sufficient to relate $\vec D(r)$ to the magnetic flux
$\phi(r)$ within a radius $r$, specifically
\begin{equation}
            \vec D(r) = {\phi(r) \hat r \over 2 \pi r} = {\phi [ 1 -
               \exp( - r^2/\sigma^2) ] \hat r \over 2 \pi r }\; .
\end{equation}
It can be shown that if the assumption of rotational symmetry is dropped,
this formula leads only to an underestimate of the effect.

We now would like to calculate the total X-ray flux from a sunspot.  The
largest sunspots we will imagine will have a penumbra of size $p =
.063R_\odot$ \cite{allen}.  This is somewhat smaller than the scale on
which our underlying pseudoscalar production changes.  It therefore seems
not unreasonable to approximate $F(E,b)$ in (\ref{counts}) by the value
at the center of the sunspot.  Numerical calculations
approximating the variation of $F(E,b)$ throughout a very large sunspot
show that this introduces at most a few percent error on the X-ray count
rate.  For a more typical sunspot, $p=.02R_\odot$, the error drops below
1\%.  With these two simplifications and (\ref{counts}), we find that the
differential X-ray flux for sunspots is given by
\begin{equation}
      {d\Phi_s \over dE} = {g^4 R_\odot F(E,b) \over 64 \pi^4 L^2}
              \int_0^p 2\pi r\,dr \left| \vec D(r) \right|^2
      = {g^4 R_\odot \phi^2 F(E,b) \over 128\pi^5L^2} \bigl[ 0.35 \bigr]
              \; ,
\label{spotcounts}
\end{equation}
Letting $g=10^{-10} {\rm GeV}^{-1}$ and optimistically setting
$\phi = 10^{23}\,$ maxwells, which implies a very large sunspot with
$p\approx .06R_\odot$ and $ B_z(0) \approx 3900\,$gauss, we find
\begin{equation}
             {d\Phi_s \over dE} = 540 \left(F(E,b) \over {\rm keV^5} \right)
                  \rm counts \, s^{-1} cm^{-2} keV^{-1} \; .
\end{equation}
The maximum value at $E=4\,$keV and $b=0$ gives about $410 \rm \, counts \,
cm^{-2}s^{-1}keV^{-1}$.

Both (\ref{counts}) and (\ref{spotcounts}) assume that pseudoscalars are
massless particles.  If they are not massless, the count rate given
above can decrease significantly.  As seen in (\ref{phase}), the
conversion of massive pseudoscalars along different points along the line of
sight will produce photons of varying phase.  For a sunspot
with $\phi= 10^{23}\,$maxwells, we find that pseudoscalars with mass
$m<10^{-5}$eV and energy $E=4\,$keV can cause a phase shift no greater than
$\pi/2$; hence, the decrease in count rate would be minimal.   However, for
masses that are much greater, incoherent contributions of photons make the
method proposed here unrealistic.   Realistic pseudoscalar masses are
either zero or typically in the range of $10^{-5}$--$10^{-2}\,$eV
\cite{pmass}.  For most of this range, the signal is reduced, probably to
invisibility.  But for masses near $10^{-5}\,$eV, there will be substantial
signal, though somewhat reduced from the signal calculated in this paper.
We will not calculate in the boundary region $m \simeq 10^{-5}\,$eV
explicitly, but will simply state that any limits obtained by this method
using a detector that sees X-rays around 4$\,$keV will apply only for masses
$m < 10^{-5}$eV.

\section{Detection of X-rays Converted from Solar Pseudoscalars}

As evident from Fig.\ 2, most of the solar X-rays arising from
pseudoscalar-photon conversion are in the energy range of 1--15$\,$keV.
At present, the best detector for finding such X-rays is the Yohkoh
satellite, which is equipped with X-ray detecting instruments for the
primary purpose of studying solar flares.  In particular, the Soft X-ray
Telescope (SXT) on Yohkoh forms X-ray images in the 0.25 to 4.0$\,$keV
range \cite{SXT}.  Because the expected signal detected by the SXT will be
concentrated around 3$\,$keV, we will only assume validity of our result
for masses $m < 7 {\times} 10^{-6}$eV.

SXT has an effective area in this energy region of a little under
1$\,$cm$^2$.  Multiplying SXT's unfiltered effective area with
(\ref{spotcounts}) and integrating through the energy range, we obtain the
count rate detected by SXT from a sunspot due to pseudoscalars.  Once again,
assuming $g=10^{-10} {\rm GeV}^{-1}$ and a very large sunspot such that
$\phi = 10^{23}\,$ maxwells, we obtain SXT's expected count rates for
various $b$, as graphed in Fig.\ 4 with the maximum count rate being
around 190$\,$counts/s at $b=0$.  (In contrast, the signal from the whole
quiet sun dipole field will be only 0.04$\,$counts/s.)

Unfortunately, the signal is dominated by background, which for such a
large sunspot would be of the order of 4100$\,$counts/s \cite{hudson}.
This does not necessarily make our signal unobservable.  A single sunspot
could be watched while it traverses across the face of the sun.  The rate
could be measured before it reaches the center of the sun, and compared to
the rate while it is exactly on center.  If the coupling $g=10^{-10}\rm
GeV^{-1}$, then there should be a four percent increase in the X-ray count
rate.  If systematic errors can be reduced below this level, then couplings
at this level can be discovered or ruled out.

If there are no systematic errors to worry about, then using a $10^3$
second exposure would allow one to see signals as low as 11$\,$counts/s at
the $5\sigma$ level.  Since the count rate scales as the fourth power of
the coupling, this allows us to limit or discover the coupling down to
$g=5{\times}10^{-11}\,\rm GeV^{-1}$.  This assumes we can reduce systematic
problems to about the $10^{-3}$ level.

Of course, there are systematic contributions.  The penumbra may contain
hot spots which produce many more X-rays than the rest of the sunspot.
These hot spots may be transient, so that they might, for example, turn on
coincidentally as the sunspot crosses the center.  This effect can be
limited by using the image to identify, isolate, and eliminate these
hotspots.

Perhaps a more difficult problem is that the viewing angle of
the sunspot changes as it crosses the face of the sun.  This changes the
apparent size of the sunspot, as well as changing the thickness of solar
atmosphere through which it is viewed.  One can argue that such effects
should be proportional to $b^2$, the square of the impact parameter, and
hence should be of the order of only a few percent or so as the sunspot
moves from $b=0.2$ to $b=0$.  Thus it does not necessarily hamper us at the
level of $g=10^{-10}\,\rm GeV^{-1}$, and by correcting for it we can
probably improve on this limit.

There are a variety of other potential systematic headaches, such as
angle-dependent detector efficiencies, evolution of sunspot as it moves,
and perhaps others.  None of these seem to be insurmountable, but they must
all be systematically addressed before the value of $g$ can be reliably
discovered or ruled out.  We believe that the ultimate limit from this
approach using the Yohkoh detector will be in the range $g=0.5$--$1 \times
10^{-10}\,\rm GeV^{-1}$.  This is the limit for mass $m <
7{\times}10^{-6}$eV.

\section{Conclusions}

We have found that searching for X-rays from the sun could limit
the photon coupling of a light pseudoscalar.  The X-ray production from the
sun's dipole moment is negligible, but with slightly optimistic assumptions
about sunspots, we can obtain a limit using Yohkoh's SXT from the umbral
plus penumbral X-ray signal of about
\begin{equation}
             g < 0.5\hbox{--}1 \times 10^{-10}\,{\rm GeV^{-1}} \; \qquad
                   \hbox{for} \qquad m < 7{\times} 10^{-6}\, \rm eV \; .
\end{equation}
This limit requires that we study the time evolution of the X-ray
production from sunspots as they cross the center of the sun, looking for a
brightening in the region $b< 0.1$.  The exact limit depends to what extent
systematic uncertainties can be eliminated.  How does this compare with
other approaches?

Raffelt and Dearborn have previously placed limits on this coupling at the
level $10^{-10}\,\rm GeV^{-1}$ by considering the effects of pseudoscalars
on the lifetime of horizontal branch stars \cite{hbstars}.  Naively, our
limit looks as good or better than theirs, but only marginally so.  In
fact, their limit applies to masses as high as $1\,$keV, and hence their
limit applies in the all-important axion case, whereas ours does not.
Furthermore, Raffelt has recently reexamined the limit from this approach,
and concluded that it can limit the coupling down to $6{\times}10^{-11}\,
\rm GeV^{-1}$ \cite{pmass}, covering nearly all of the range of our
detection.

In a previous paper \cite{carlson}, one of us considered what limit had
been set or could be set by looking for X-rays from stars,
particularly the nearby supergiant $\alpha$-Ori.  A limit of $g < 2.5
{\times}10^{-11}\,\rm GeV^{-1}$ was set by reexamining old data, and it was
pointed out that this limit could be improved to the level of
$10^{-11}\,\rm GeV^{-1}$ if a dedicated observation were made.  This is
better than we believe the solar approach can achieve, but it applies only
to masses below about $10^{-10}$eV.

Combining the other two limits, we see that Yohkoh has a small window of
opportunity for this method for masses in the range $10^{-10}$--$7{\times}
10^{-6}$eV, if the coupling is very close to $5{\times}10^{-11}\,{\rm
GeV}^{-1}$.  How likely is this value?  It is
not favored by any model of which these authors are aware.  For such a
small coupling, eliminating systematic errors is crucial, and any
``discovery'' of a signal would be met with justifiable skepticism.

We therefore conclude that the approach we have proposed here is of only
marginal value for investigating new portions of parameter space.  On the
other hand, more information is better.  In the field of
astrophysics, one never knows when some limit will be called into question,
and it is always good to have redundant approaches to particle physics
limits.  Redundant approaches are as close as we can come to repeatability
of experiments in astro-particle physics, so we should seize every
opportunity we can find.  Who knows?  Maybe this approach will find
evidence for pseudoscalars, or perhaps just lead to a better understanding
of X-rays from sunspots.

\acknowledgements
We would like to thank H.\ Hudson for providing extremely helpful
information concerning Yohkoh's SXT.  We would also like to thank G.\ Field
for his helpful input to this project.  Additional thanks go to J.\
Bookbinder and L.\ Golub for helpful comments on solar X-ray detection.
This research was supported by the National Science Foundation under
contract {\tt NSF-PHY-92-18167}, and also by a Sloan Foundation Fellowship.
L.S. Tseng was supported by a grant from the Harvard University Physics
Department Undergraduate Research Fund.

\bigskip\bigskip

\centerline{{\bf Figure Captions}}

\def\figg#1{\bigskip\bigskip\hang\noindent {\bf Fig.\ #1.}}

\figg1 The temperature (solid line) and $\xi^2$ (dashed line) as functions
of radius $r$ based on the model of Bahcall and Ulrich \cite{model}.

\figg2 The function $F(E,b)$ for $b$ from $b=0.0$ (top curve) to $b=0.3$
(bottom curve).  The solid lines are at intervals of $\Delta b=0.1$, and
the dashed lines are at intervals of $\Delta b=0.02$.

\figg3 The expected X-ray signal due to pseudoscalar-photon conversion in
the sun's general dipole field, taking $g=10^{-10}\,\rm GeV^{-1}$ and
$B_p=1\,$gauss.

\figg4 The expected X-ray count rate detected by Yohkoh's unfiltered SXT
due to pseudoscalar-photon conversion in the magnetic field of a
sunspot, taking $g=10^{-10}\,\rm GeV^{-1}$ and $\phi = 10^{23}\,$maxwells.
The sunspot's center defines the impact parameter $bR_\odot$.

\vfill\eject
\end{document}